%% file: wsdm2018-xinyi.tex
\documentclass[format=sigconf,letter]{acmart}
\renewcommand\footnotetextcopyrightpermission[1]{} % removes footnote with conference information in first column
\if0
\usepackage[
  pass,% keep layout unchanged
]{geometry}
\fi
\setcopyright{acmcopyright}

\usepackage{booktabs,times}
\usepackage{caption}
\usepackage{subcaption}
\usepackage{tikz}
\usetikzlibrary{fit,positioning,matrix,chains,positioning,arrows}
\usepackage{ctable} 
\usepackage{color}
\usepackage{url}
\usepackage{mathtools}
\usepackage{setspace}
\usepackage{latexsym}
\usepackage{float}
\usepackage{amsmath,amssymb}
\usepackage{paralist}
\usepackage{enumitem}
\setlist{nosep}
\usepackage{verbatim}
\usepackage{algorithm}  
\usepackage{algorithmic} 
\usepackage{textcomp}
\usepackage[skip=0pt]{caption}
\newcommand{\xinyi}[1]{\textcolor{magenta}{(\textbf{XL:} #1)}}

\newcommand{\ms}[1]{\textcolor{red}{(\textbf{MS:} #1)}}

\title{Characterizing Reading Time on Enterprise Emails}

\author{Xinyi Li}
\affiliation{%
  \institution{University of Amsterdam}
  \city{Amsterdam} 
  \country{The Netherlands} 
}
\email{lixinyimichael@gmail.com}

\author{Chia-Jung Lee}
\affiliation{%
  \institution{Microsoft}
  \city{Redmond} 
  \country{USA} 
}
\email{cjlee@microsoft.com}

\author{Milad Shokouhi}
\affiliation{%
  \institution{Microsoft}
  \city{Redmond} 
  \country{USA} 
}
\email{milads@microsoft.com}

\author{Susan Dumais}
\affiliation{%
  \institution{Microsoft Research}
  \city{Redmond} 
  \country{USA} 
}
\email{sdumais@microsoft.com}

\begin{document}

\begin{abstract}
Email is an integral part of people's work and life, enabling them to perform activities such as communicating, searching, managing tasks and storing information. Modern email clients take a step forward and help improve users' productivity by automatically creating reminders, tasks or responses. 
The act of \textit{reading} is arguably the only activity that is in common in most -- if not all -- of the interactions that users have with their emails.

In this paper, we characterize how users read their enterprise emails, and reveal the various contextual factors that impact reading time. Our approach starts with a reading time analysis based on the reading events from a major email platform, followed by a user study to provide explanations for some discoveries.
We identify multiple temporal and user contextual factors that are correlated with reading time. For instance, email reading time is correlated with user devices: on desktop reading time increases through the morning and peaks at noon but on mobile it increases through the evening till midnight. The reading time is also negatively correlated with the screen size.

We have established the connection between user status and reading time: users spend more time reading emails when they have fewer meetings and busy hours during the day. In addition, we find that users also reread emails across devices. Among the cross-device reading events, 76\% of reread emails are first visited on mobile and then on desktop. Overall, our study is the first to characterize enterprise email reading time on a very large scale. The findings provide insights to develop better metrics and user models for understanding and improving email interactions. 
\end{abstract}

\keywords{Email reading time; Enterprise emails; Cross-device email reading}

\maketitle
\thispagestyle{empty}
%%Introduction
\input{wsdm2018-xinyi-1}
%%Related work
\input{wsdm2018-xinyi-2}
%%Method
\input{wsdm2018-xinyi-3}

%%Overview
\input{wsdm2018-xinyi-4}

\input{wsdm2018-xinyi-5}
\input{wsdm2018-xinyi-6}

%%User behavior
\input{wsdm2018-xinyi-7}

%User study
\input{wsdm2018-xinyi-8}

%%Conclusion
\input{wsdm2018-xinyi-9}
\if0
\begin{spacing}{1}
\small\smallskip\noindent\textbf{Acknowledgments.}
This research is supported by Microsoft.
\end{spacing}
\fi

\bibliographystyle{abbrvnatnourl}
\bibliography{refs}  

\end{document}

%% file: wsdm2018-xinyi-1.tex
% !TEX root = ./wsdm2018-xinyi.tex

\section{Introduction}
\label{sec:intro}

Emails are one of the most important channels for communication~\citep{pew2014}. Over the past two decades, the nature of web emails has significantly evolved and influenced user behavior accordingly~\cite{Maarek:2016:MNF:2835776.2835847}. Email usage has become much more diverse including task management, meeting coordination and personal archiving and retrieval. 
% and the majority of email messages in \textit{consumer}\footnote{Not to be confused with enterprise email accounts that often contain work-related emails.} accounts are now generated by machines~\cite{Ailon2013}.  %\sd{the Yahoo work is for personal accounts; not sure if this is also true for enterprise accounts} 
The high demand for intelligent email systems fostered related research in many areas such as email search~\cite{ai2017,Carmel2015}, information organizing with folders or tags~\cite{Grbovic:2014:MFY:2661829.2662018,Koren:2011:ATE:2020408.2020560}, and predicting user actions of replying or deleting~\cite{Dabbish:2005:UEU:1054972.1055068,DiCastro2016,kooti2015evolution,Yang:2017:CPE:3077136.3080782}.
Although prior work has provided in-depth analyses and solutions for specific applications, the fundamental understanding of how users interact with email clients remains somewhat unclear. For example, questions such as how and when people read emails, how long they spend doing so, and what factors influence reading are not well understood.

The goal of this study is to characterize and present a comprehensive view on how users \textit{read} their emails and how their \textit{reading time} is affected by various contextual cues.

The reading activity is embedded in most user-email interactions across diverse applications ranging from retrieving information to automatic email prioritization. 
We argue that understanding email reading time lays the ground work for understanding user satisfaction, as it paves the way to estimating how a user's focus is spent.
Capturing user reading time also helps reasoning about how email clients can be improved. 
Properly characterizing reading time, however, is a very challenging task. In today's environment, email clients are built with rich functionalities and using multi-devices by a single user is common. Even with access to large-scale logs, it requires careful examinations on data selection and interpretations to deliver meaningful analysis. 

In this paper, we provide a quantitative analysis of enterprise emails from the web and mobile clients of a popular email web service. We start by introducing a method to approximate reading time, that can be applied on millions of emails (Section~\ref{sec:method}). Then, we uncover how reading time is affected by various contextual factors in Section~\ref{sec:observations}. We delve into temporal factors, user contextual factors and a very common reading behavior -- rereading. To complement the results based on the log analysis, we also conduct a user study to look for the causes behind some interesting observations (Section~\ref{sec:userStudy}).

Our findings indicate that reading behavior differs significantly on desktop versus on mobile devices. While the majority of emails are read within 10 seconds on both clients, the distribution of reading time on desktop exhibits a heavier tail than on mobile. Further, we find that desktop and mobile users have different temporal patterns: on desktop the reading time increases through the morning whereas on mobile it increases from the evening till midnight. Email types are also correlated with reading time: e.g. on restaurant and hotel related emails, users spend longer time during weekends compared to weekdays. The average time spent on reading emails is dependent on user status as well. For example, users spend less time reading when their calendar status is ``out of office.'' They also read fewer emails within shorter time when they have more meetings or are busier in a day. We find different reading patterns in cross-device reading events: for instance, when users switch from mobile to desktop, email reading time tends to increase; when they switch vice versa, however, reading time tends to decrease. Last but not the least, our user study sheds light on on why certain behaviors occur.

To the best of our knowledge, this study is the first of its kind to uncover how users spend time reading emails through a large-scale analysis. The findings enrich the understanding of email reading behavior, and benefit research and applications in this field. 
For instance, correct interpretation of reading time would be essential for determining the importance of an email for email prioritization features\footnote{Examples include, Outlook Focused Inbox, or Gmail Priority Inbox.}, or its relevance in information seeking scenarios. 
Since reading time differs by contexts, the same amount of time spent on a human-authored email and a machine-generated email may mean different degrees of relevance. 

%% file: wsdm2018-xinyi-2.tex
% !TEX root = ./wsdm2018-xinyi.tex

\section{Related work}
\label{sec:relatedWork}

A rich spectrum of studies have been conducted on users' interactions with email clients. In this section, we provide an overview of the most related work to our study. 

\textit{Email overload and prioritization}. 
Information overload was identified in the early years as one of the critical issues for email users \cite{Dabbish:2006:EOW:1180875.1180941,Whittaker:1996:EOE:238386.238530} and still is prominent in current email systems \cite{Grevet2014}. Beyond spam filtering techniques \cite{Dasgupta:2011:EES:1935826.1935929}, \citet{Yoo:2011:MPE:2063576.2063683} focused on modeling personal email prioritization to make more critical information surface to the users. \citet{Wainer:2011:IOT:1978942.1979456} examined how top-level cues such as message importance, subject line specificity, workload and personal utility influence users' attention to emails. \citet{aberdeen2010learning} introduced a scalable learning system to classify each email as important or not important for Gmail Priority Inbox, where the classification threshold is personalized per user.

\textit{Email search}. In addition to the above proactive scenarios, users interact with and rely on search to retrieve relevant information  or to organize emails. \citet{kruschwitz2017searching} demonstrate that email search is an essential part of the information seeking behavior in enterprises. \citet{ai2017} have examined the search behavior on email systems. They characterized multiple search strategies and intents, and identified important behavioral differences from web search such as re-finding. 
\citet{Horovitz:2017} proposed an auto-completion feature for email search, were suggestions are extracted from personal mailbox content in addition to query logs from similar users. 
\citet{Narang2017} investigated general email activities and search activities. They found that search strategies are correlated with mail box properties as well as organizing strategies. \citet{Kim:2017:UMS:3077136.3080837} studied email search success by popping up an in-situ survey when a search session is finished to collect feedback. The results showed that generative Markov models can predict the session-level success of email search better than discriminative models.  Along the line of searching personal information, \citet{Dumais2016} examined in detail users' reusing behavior and established systems that assist users to find items such as emails and documents that users have seen before. \citet{Cecchinato:2016:FEM:2858036.2858473} investigated different finding strategies on desktop versus mobile devices, and work versus personal accounts via a diary study.   Additional efforts \cite{Carmel2015,Carmel:2017:PRR:3038912.3052659,Carmel:2017:DMS:3038912.3052658,Kuzi:2017:QEE:3077136.3080660,Ramarao:2016:IRE:2911451.2911458} have also been laid on improving ordering accuracy for better search experiences. 

\textit{Email interactions}. Users tend to perform a variety of actions in email clients. ~\citet{DiCastro2016} conducted large-scale log analysis for predicting users' actions of reading, replying, forwarding and deleting after receiving an email. ~\citet{Yang:2017:CPE:3077136.3080782} focused on predicting the reply action and studied the impact of factors such as email metadata, historical interaction features and temporal features. \citet{Dabbish:2005:UEU:1054972.1055068} studied the decision rules people  choose to
reply to email messages, or to save or delete them through a survey.

\textit{Folders and tags}. Email systems not only provide a communication channel but users often manage their personal information by taking actions such as archiving, tagging or foldering. Earlier studies tackled the task of auto-foldering for individuals where the goal is to classify each email into a user defined folder \cite{bekkerman2004UMass,dredze2008IUI,tam2012ECIR}. More recently, \citet{Grbovic:2014:MFY:2661829.2662018} proposed to address the sparsity problem arising from the earlier personalized approaches by inferring common topics across a large number of users as target folders. \citet{Koren:2011:ATE:2020408.2020560} associated an appropriate semantic tag with a given email by leveraging user folders. \citet{wendt2016hierarchical} proposed a hierarchical label propagation model to automatically classify machine generated emails.

\textit{Email intelligence}. Current email clients aim to help users save time and increase productivity. \citet{Kannan:2016:SRA:2939672.2939801} investigated an end-to-end method for automatically generating short email responses as an effort to save users' keystrokes. \citet{Ailon2013} proposed a method to automatically threading emails for better understanding using causality relationship. Email summarization \cite{Carenini:2007:SEC:1242572.1242586,Muresan:2001:CLM:1117822.1117837} has been studied as a promising way to solve the problem of accessing an increasing number of emails possibly on small mobile devices. 

While prior work studied extensively from different perspectives how users interact with email systems, their focuses were centered around specific scenarios such as search. The goal of this paper is to present a horizontal, generic view on users' interactions with emails in terms of reading, which is the primary action users take regardless of which application they are currently using. Not only do we study in detail the relations between reading time and a variety of properties, but we contrast the reading behavior on desktop and mobile devices over a large number of real users.

In their highly cited work on \textit{Theory of Reading}, \citet{just1980theory} argue that reading time depends on text, topic and the user familiarity with both. Almost four decades later, we reassess some aspects of their theory on user interactions with modern emails.

%% file: wsdm2018-xinyi-3.tex
% !TEX root = ./wsdm2018-xinyi.tex

\begin{figure*}[t]
        \centering  
         \includegraphics[width=4.5in]{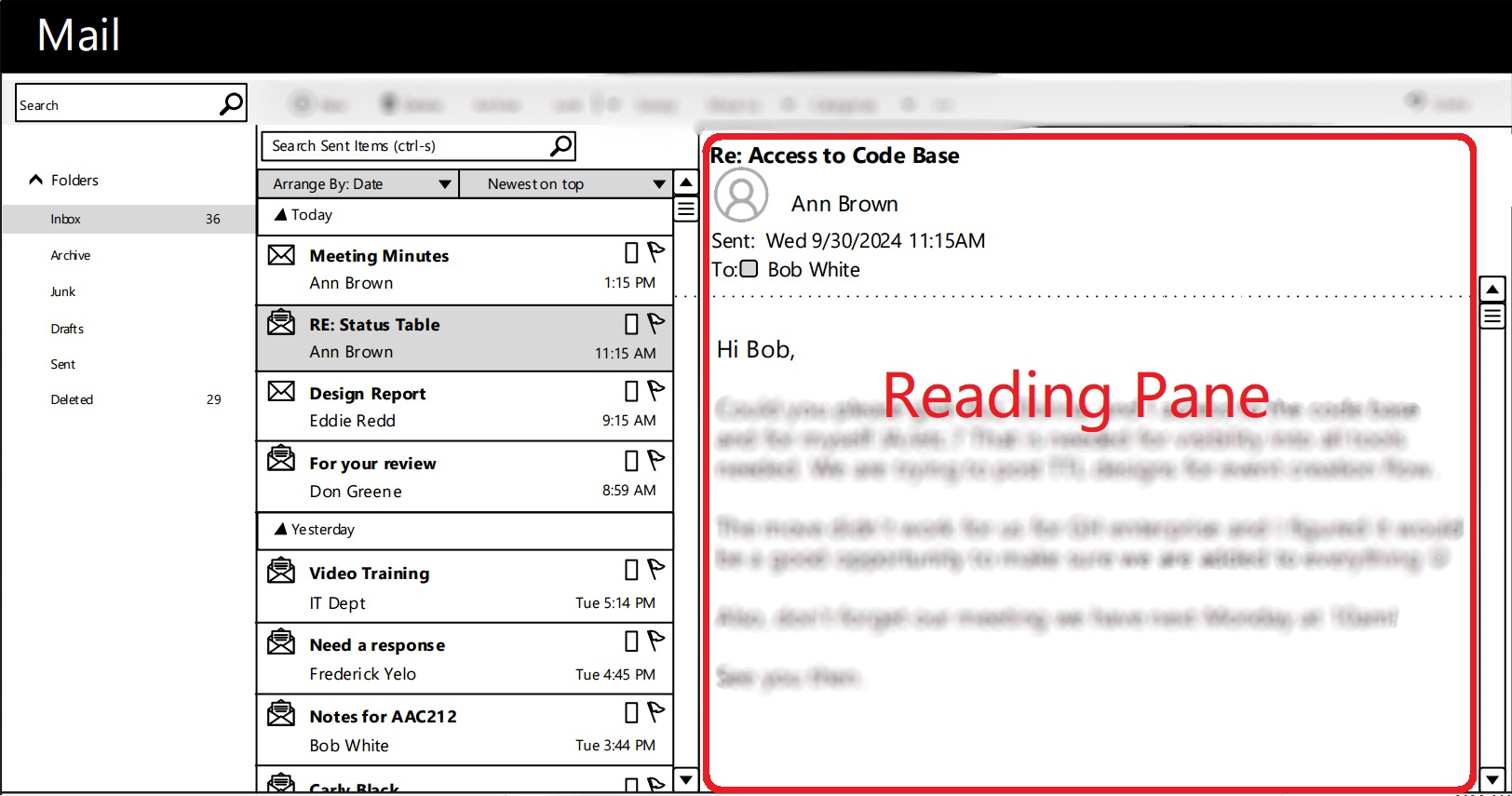}\quad\quad\quad
         \includegraphics[width=1.45in]{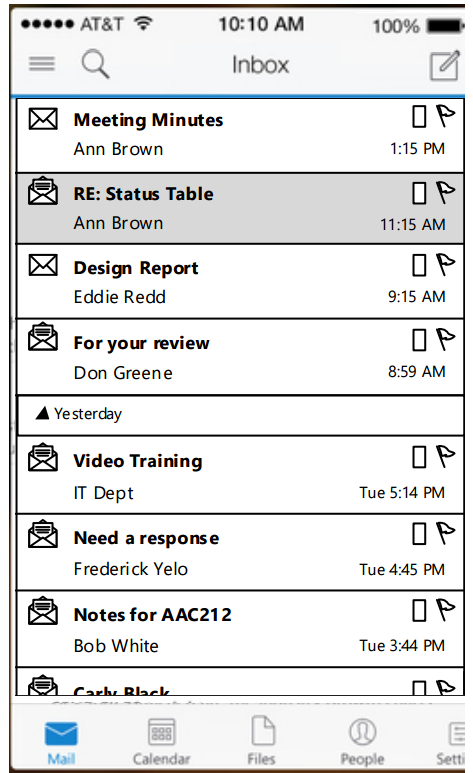}
         \caption{The web interface (left) and the mobile app interface (right) for our email clients. The reading time on desktop is computed with respect to the time each message appears in the reading pane (red box). The reading time is computed from the moment an email is clicked on until the the user hits back (available on mobile only), clicks on the next listed email (available desktop only), switches to compose mode by clicking on reply/forward, or closes the app (or browser).}
\label{fig:owa-screenshots}
\end{figure*}

\section{Measuring Reading Time}
\label{sec:method}
Measuring reading time accurately is challenging. Eye-tracking tools can be used to track the users' gaze, but deploying them over large numbers of users is non-trivial due to privacy concerns, costs and technical limitations around calibration. 
We rely on user interaction logs of a large commercial email provider to study the reading time indirectly by measuring the time between opening and closing an email. 
Relying on interaction logs allows us to test our hypotheses over large sets of users at reasonable costs and with minimal intrusion. 
However, our data-driven approach is limited to what is already captured in the logs, and is not free of issues.
For instance, people might be multi-tasking -- they might have the email opened but are focusing on a different task in a different window. Furthermore, a logged \textit{open} action on an email followed by a logged \textit{close} action does not always imply that the email is \textit{read} (e.g., the user might be triaging emails quickly, deleting emails as soon as they are displayed on screen). 

In our analysis, we use the best possible signals in the logs to get a close approximation of the reading time. We define reading time as the duration between the two paired signals -- the start of email reading pane which loads the content of an email into the reading zone and the end of email reading pane which records the closing of that pane, as it forms a consecutive \textit{reading event}. To minimize potential impacts caused by the above issues, we ignore samples with reading time shorter than one second. Reading events on threads (20.5\%) are removed since they are more conversational in nature and complex to track.\footnote{The dwell time on each email of a thread is dependent on the size of the screen, scrolled position of the pane and other factors.} We also only study users who read at least one email per weekday so as to focus on normal traffic and avoid random noises.
\paragraph{Data}

Our experimental data is sampled from enterprise emails over a two-week period from May 6th to May 20th 2017. We enforce the above filtering rules when collecting the data. Beyond this, we sample the data randomly to minimize potential biases towards specific demographics or enterprises. For simplicity, we refer to this dataset as desktop client dataset. In total, this sample contains 1,065,192 users, 69,625,386 unique emails\footnote{One email that is sent to multiple recipients is counted as one.} and 141,013,412 reading events (i.e., an average of 132 reading events per person) from tens of thousands of enterprises. From this set, we further select users who also use the iOS app over the same period and collect their corresponding usage from the mobile logs, which is referred to as the mobile dataset. This gives us 83,002 users with 5,911,107 unique emails and 10,267,188 reading events (an average of 124 reading events per user). By collecting email usage patterns from both desktop and mobile clients, we are able to study in-depth cross-device reading behavior. In addition to the two-week window of data, we also collect another two-week period data prior to this period from the same set of users. This ``history'' data is used to capture rereading behavior if any.
\paragraph{Desktop (web) client}
An anonymized version of the user interface of the web email client is shown in Figure~\ref{fig:owa-screenshots} (left).
The interface supports users to manage their emails effectively on web browsers. We find that nearly all the usage data logged from this portal comes from desktop/laptop users, which is why we refer to it as desktop client throughout the paper. On mobile phones, people tend to use a mobile email client (app), as described later. To read an email on our desktop client, users have to first select it from the email list by clicking on it.

Once an email is selected from the list, its corresponding content will show up instantly on the \textit{reading pane} on the right side of the email list. 
As mentioned, we use the time gap between when a message appears in the reading pane and when it is replaced with another message to approximate reading time. 

As a sanity check, we validate this method by first performing various actions on the client by ourselves and video-record everything, and then checking the corresponding logs recorded by the system. We find that for majority occasions our email reading time can be reflected by the time gap between reading pane's opening and closing. However, when we quickly navigate through emails in the email list by pressing arrow keys or clicking, a very short reading time (such as hundreds of milliseconds, but no more than one second) is recorded by the system. Given the very short time, we assume that it is unlikely for other users to read the email as well. Therefore we set a one second threshold on the reading time in order to filter out these unlikely reading events. %\ms{we should probably mention how much data we are throwing away this way. Also, can we find a reference that supports this threshold. There should be previous work on minimum time required for people to look at something to be able to read/understand/describe it.} \xinyi{didn't find such a reference}
\paragraph*{Mobile client (app)}
The right screen-shot in Figure~\ref{fig:owa-screenshots} depicts the user interface on the iOS mobile application which is the source of the mobile logs. 
Users can click into an email by tapping on an email snippet in the list display. A reading pane with the email content will show up that fills the display area of the application. The reading time is the interval between tapping an email and hitting the exit/back button or quitting the app. 

Due to data sensitivity, normalized reading time is used in some analyses instead of absolute time. Time is turned to logarithmic form and then min-max normalized to avoid showing absolute time.

%% file: wsdm2018-xinyi-4.tex
% !TEX root = ./wsdm2018-xinyi.tex

\begin{figure}[tp]
        \centering  
         \includegraphics[width=\columnwidth]			
         {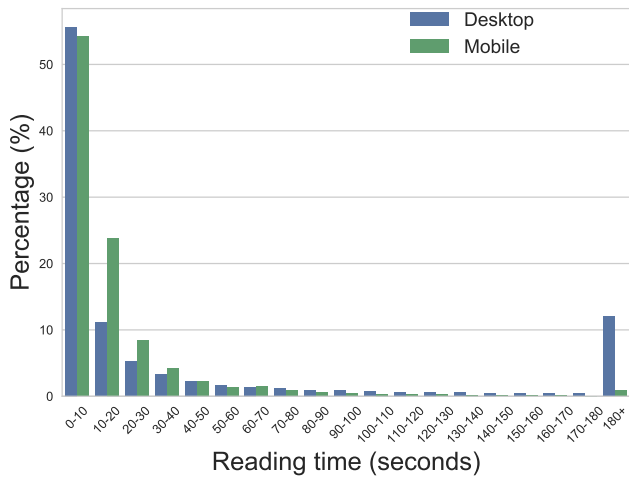}
         \caption{Reading time distributions in desktop and mobile platform. Time is binned by every 10 seconds.}
         \label{fig:overview}
\end{figure}

\if0
\begin{figure*}[tp]
        \centering  
         \includegraphics[width=\textwidth]{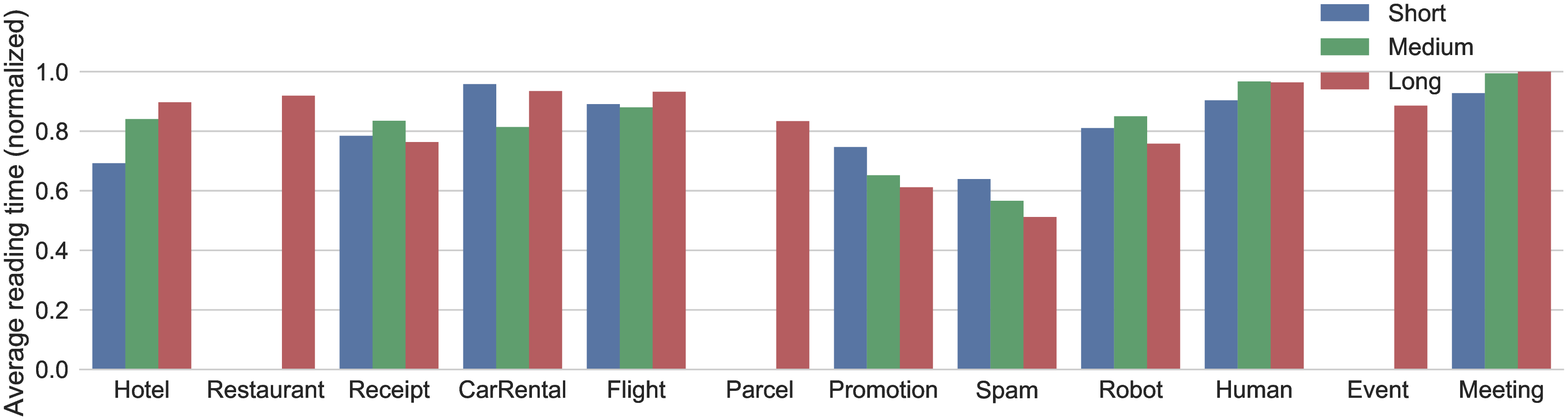}
         \caption{Average reading time by different length segmentation and email types. A proprietary multi-label text classifier has been used to classify emails into different groups based on the content of their body.}
         %\ms{Should we change this to a stackbar, where each class, gets one stackbar? Also, it would be fine to drop events with a single class from the plot and explain why they are excluded in the text and caption.}} 
         \label{fig:labelTime}
\end{figure*}
\fi

%\section{Distribution of reading time}
\section{Reading time and contextual factors}
\label{sec:observations}
In this section we first provide a brief overview of reading time, then delve into various contextual factors that impact reading time.

\subsection{Reading time overview}
\label{sec:overview}
The overall distributions of reading time on desktop and mobile are presented in Figure~\ref{fig:overview}. In both datasets, more than half of the reading events happen in less than 10 seconds (55.6\% on desktop vs. 54.2\% on mobile). On mobile, about 25\% of emails are read in 10-20 seconds. In comparison on desktop, only about 11\% of emails fall into that range.

Interestingly, the reading time distribution on desktop has a much longer tail compared to mobile. On desktop 12.0\% of reading events last longer than 180 seconds (3 minutes) which we suspect can cover many cases where the users leave an email opened on the screen while paying attention to something else -- potentially not even being at their desk. The longer tail can also be explained by the fact that spending longer time on reading could be relatively easier on larger desktop screens.

Using proprietary email classifiers, we can put email into various categories. Not surprisingly, we find that human emails have longer reading time than robot emails, with promotional and spam emails having the shortest reading time.\footnote{Our classifiers follow a semantic taxonomy where emails are first grouped into those sent by human (\textit{human}), those sent by machines (\textit{robot}) and \textit{spam}. Human and robot classifiers are exclusive and exhaustive, while the spam classifier is independently built to output confidence scores indicating how likely an email is deemed as spam. Next, for emails that are classified as robot, ngram-based classifiers are established for identifying different intents from the emails. These include classifiers that identify travel information (\textit{hotel}, \textit{car rental}, \textit{flight}), classifiers that identity reservations for food (\textit{restaurant}) or concerts/festivals (\textit{event}), classifiers that identify your purchases (\textit{receipt}) with tracking information (\textit{parcel}), and finally any coupon codes if available (\textit{promotion}). For human emails, a rule-based classifier that identifies intents asking for meetings (\textit{meeting}) is also included for our analysis. In total, 96.1\% of the emails have been classified by the system, which provides us a representative sample for comparing the distribution of email types.}

%% file: wsdm2018-xinyi-6.tex
\subsection{Temporal factors}
\label{sec:temporal}
\begin{figure}[tp]
	\centering
        	\includegraphics[width=\linewidth]{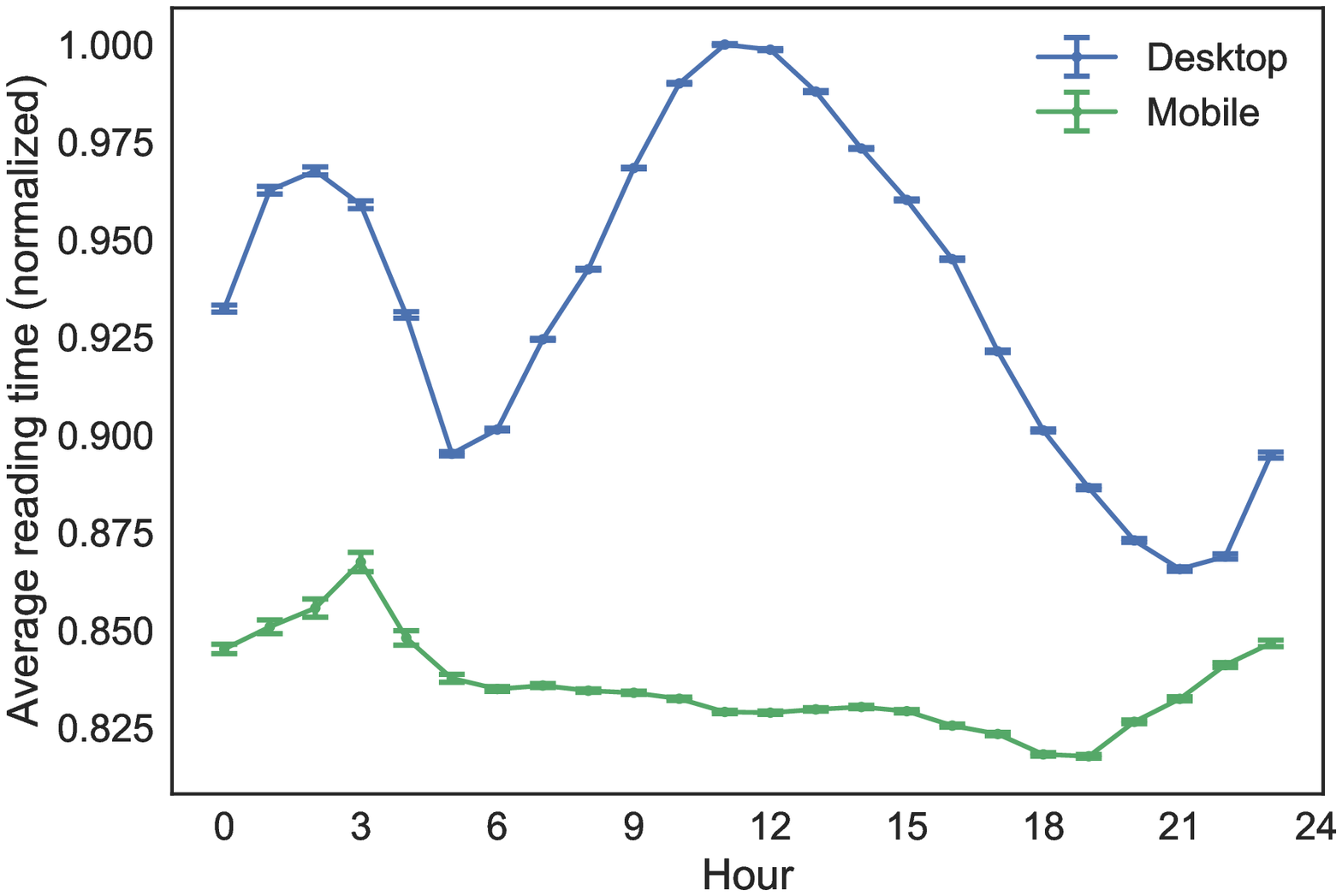}    
	\caption{Average email reading time across different hours on desktop (top) and mobile (bottom).}
	\label{fig:hour}
\end{figure}

In this section we study how reading time is affected by various temporal factors. 
To begin with, we investigate how the average time users spend reading emails varies depending on the time of day and the day of the week.
Figure~\ref{fig:hour} illustrates the average email reading time in different hours of the day.\footnote{We calibrate the calculation according to users' local time zones.} 
It can be seen that average time spent reading emails on desktops increases through morning time and peaks around noon, and then decreases through the afternoon and the evening. However, the reading time on mobile is drastically different and increases from around 7PM up until 2AM next day, while it decreases through most of the afternoon.

Moving on, we find that for both desktop and mobile, reading time on weekdays is higher than that on weekends. On both datasets, the weekday pattern is fairly stable with minor changes;
for desktop on weekdays, reading time is the lowest on Monday and highest on Friday. On mobile, the reading time slightly peaks on Wednesday and is the lowest on Monday. We omit the visual presentation of details for brevity. 

\if0
\begin{figure}[h]
	%\centering
        \begin{subfigure}[b]{0.5\columnwidth}
        \centering
        		\includegraphics[width=\linewidth]{figs/owa_readTime_by_day.eps}
        		\caption{desktop. Average time is log time in seconds.}
       		\label{fig:owaDay}
        \end{subfigure}%
        \hfill
       \begin{subfigure}[b]{0.5\columnwidth}
        \centering
                \includegraphics[width=\linewidth]{figs/mobile_readTime_by_day.eps}
                \caption{Mobile. Average time is calculated from binned time (seconds).}
                \label{fig:mobileDay}
        \end{subfigure}
    \caption{Reading time on different days in a week.} 
    \label{fig:day}
\end{figure}

\fi

\begin{figure}[tp]
	\centering
	\includegraphics[width=\linewidth]{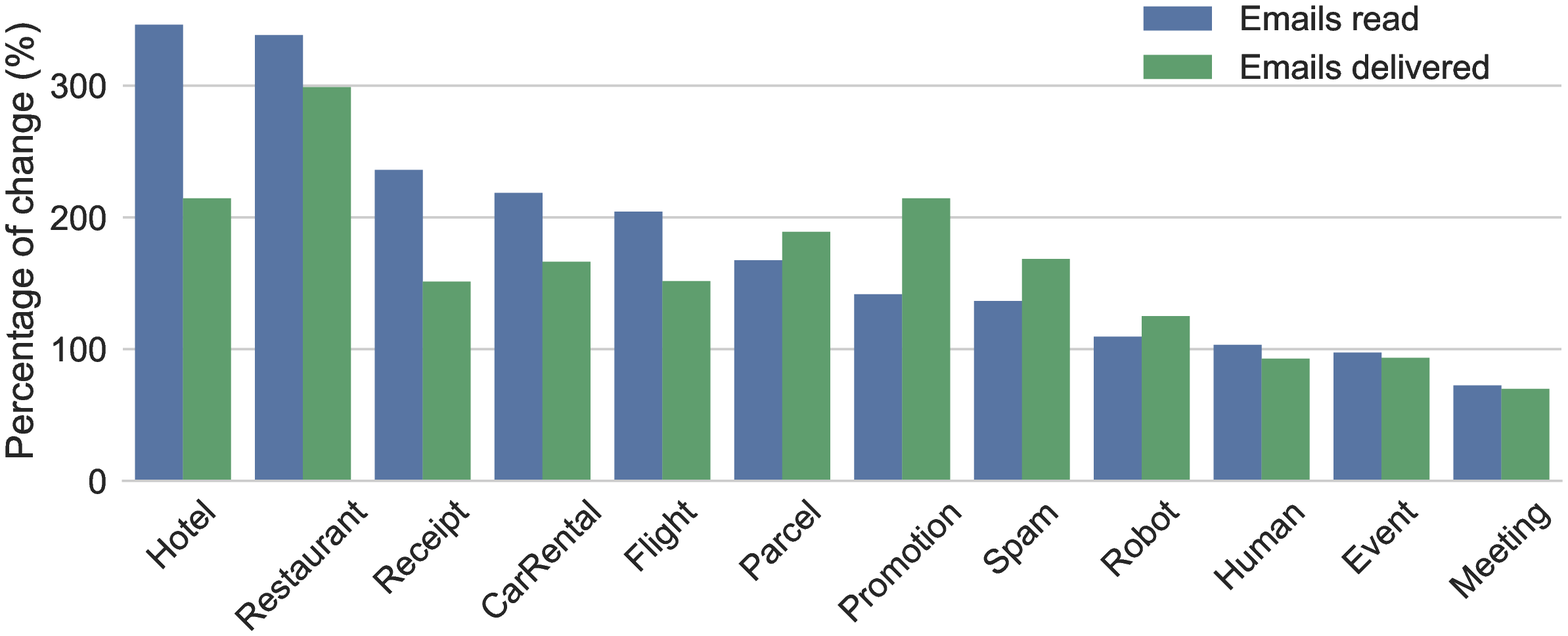}
	\caption{The percentage of changes per email type on weekends vs. weekdays. The bars are computed by dividing the frequency of emails received/read in each category over the weekend by the weekday traffic (times 100).} %and reading time in weekdays and weekends.
    \label{fig:timeComparePct}
\end{figure}

We also compare the type of emails that are typically received and read by users between weekends and weekdays, in order to find out if there is any difference among email types. The green bars in Figure \ref{fig:timeComparePct} represent the magnitude of change in percentages among emails received in each category, and are computed by dividing the number of weekend emails by those received during weekdays. For instance a roughly 100\% increase in the number of hotel-delivered emails suggests that people are almost twice as likely to receive such emails over the weekends. We also compare, how often emails from different categories are read on weekends versus weekdays. The blue bars in Figure \ref{fig:timeComparePct} -- computed in a similar fashion but based on read statistics -- confirm that the types of emails read by the users over the weekend are mostly consistent with what they receive. However, they also highlight a few exceptions where reading rates deviate from what would be expected based on delivery statistics. For instance, hotel-related emails are almost 3.5 times more likely to be read during weekends, despite the fact the number of hotel-related emails delivered only grows by a factor of 2. By contrast, spam and promotional emails are substantially less likely to be read on weekends versus weekdays.

As a brief summary, through the analyses on temporal factors, we find that the temporal pattern of reading time is not only correlated with the hour and day of the week, but also devices and certain email categories.

\subsection{User contextual factors}
\label{sec:cognition}
\begin{figure}[tp]
	\centering
	\includegraphics[width=\linewidth]{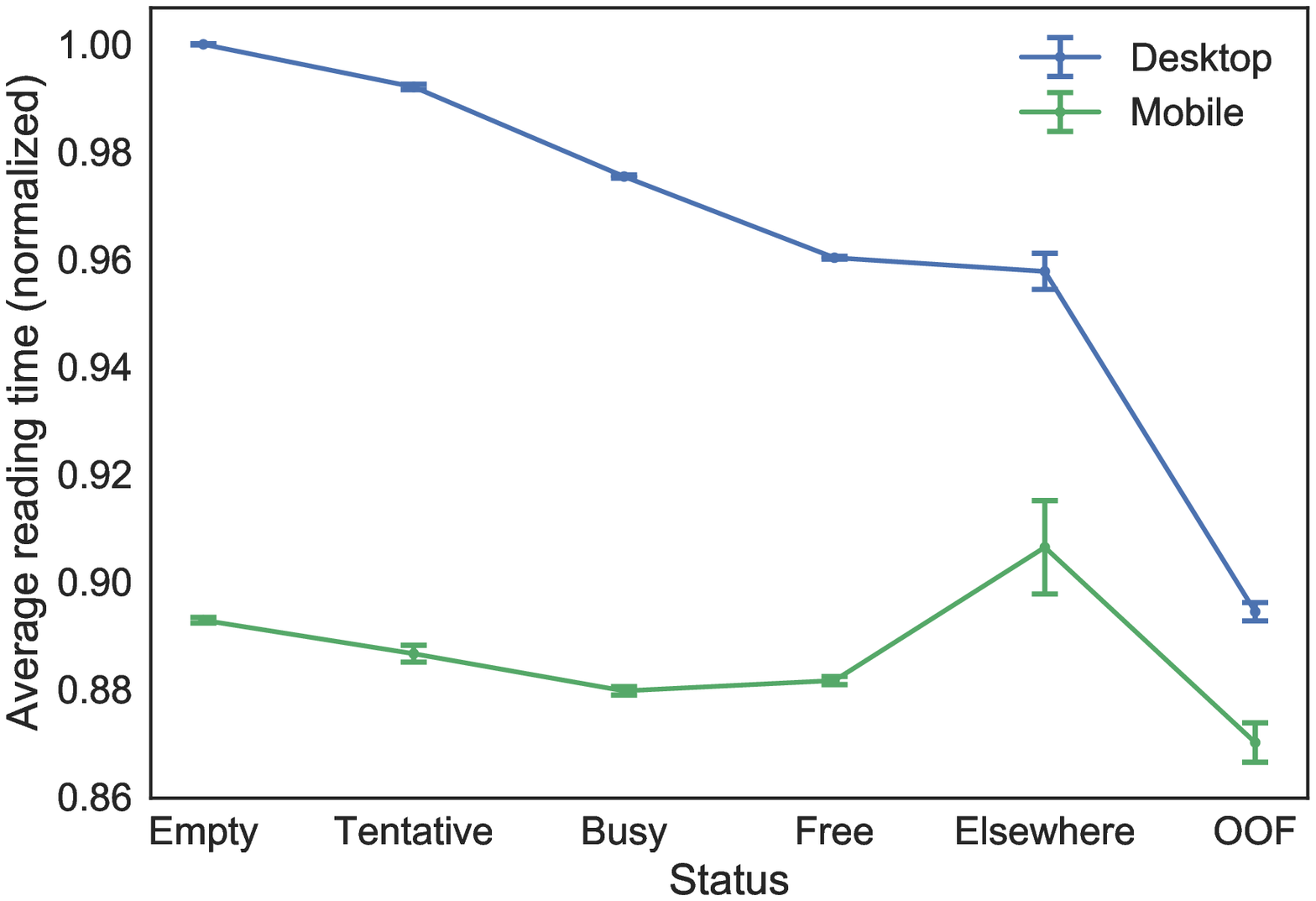}
	\caption{Reading time by different calendar status on desktop (top) and mobile (bottom).} 
	\label{fig:status}
\end{figure}

In this section, we investigate how user contextual factors could potentially affect the reading time. Specifically, we examine calendar status, fatigue and user device. 
%We use the cognitive load to refer to how busy or engaged a user may be. The cognitive load during reading can be attributed to different factors
\paragraph*{Calendar status}
\label{sec:calendar}
While we do not have direct access to the user status of our users, we can use their calendar status -- which can be mined from the calendar app associated with their email client -- as proxy to hypothesize about their user status.
The status classes include tentative, busy, free, elsewhere (working elsewhere), and OOF (out of office). Note that an empty status means nothing is on the calendar in that period, while ``free'' is a status that a user explicitly puts on the calendar and hence it suggests ``availability''. % \ms{What's the difference between empty and free}
%We group the email reading events by different statuses and examine the reading time.
The average reading time provided in Figure~\ref{fig:status} shows that the reading patterns can be affected by the \textit{calendar pressure} of the user in both platforms. On desktop, reading time is the longest when there is nothing on the calendar and the user is free. On mobile, the peak occurs at working elsewhere,\footnote{We do not have a strong explanation for long reading times on mobile when the user status is \textit{elsewhere}. The high variance (indicated by large error bars) suggests that this is not a frequent/consistent event. We can only conjecture that the spike might have been caused by users that had to read the emails they normally read on desktop at work, on their mobile devices.
}, which is the second lowest on desktop. Users tend to spend more time on an email on average, when they have nothing specific on their calendar.  They spend the shortest time when they are out of office, perhaps reading emails fast enough mainly to cherry pick the key points.

\begin{figure}[tp]
	\centering
        		\includegraphics[width=\linewidth]{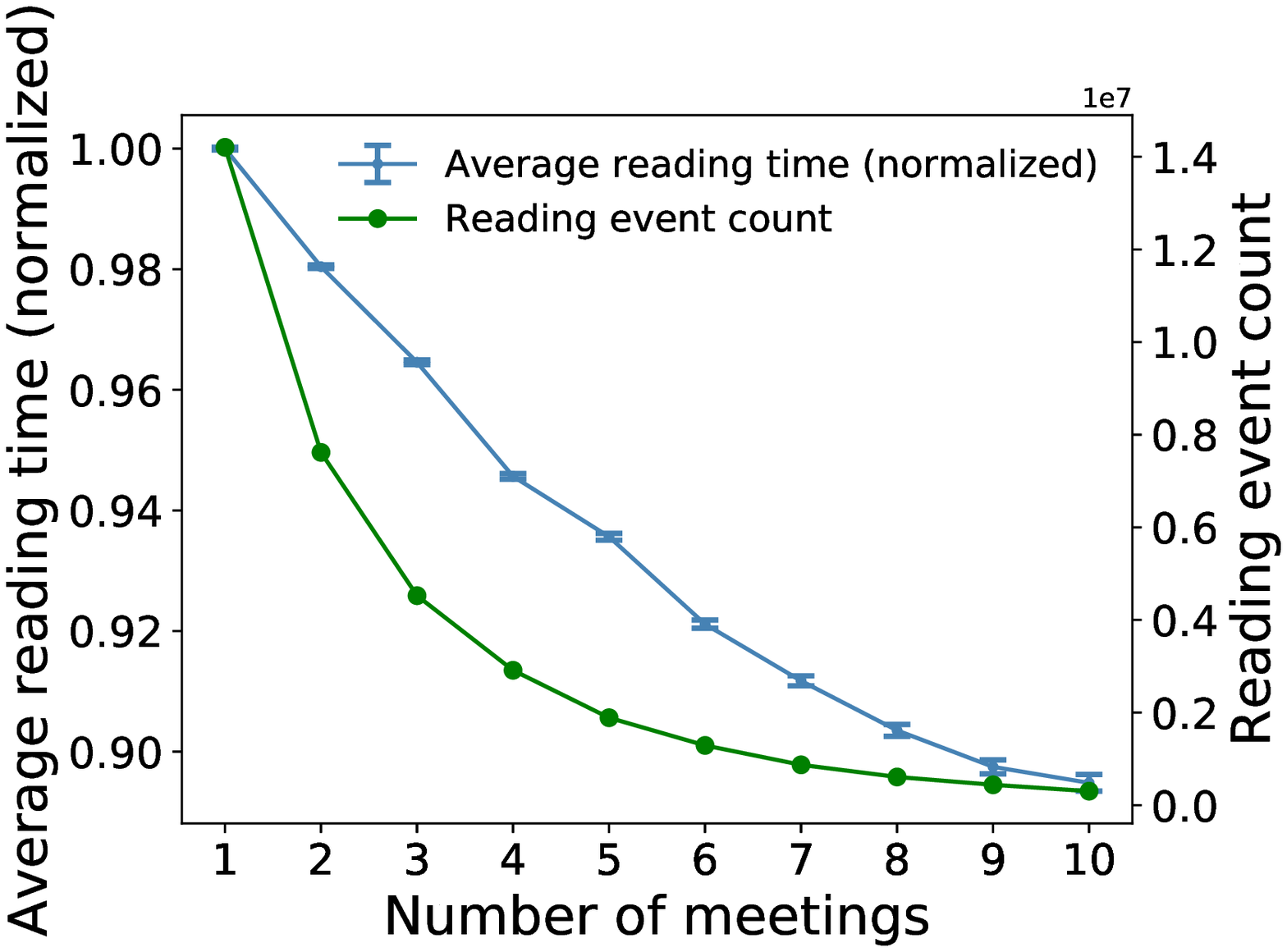}
                \includegraphics[width=\linewidth]{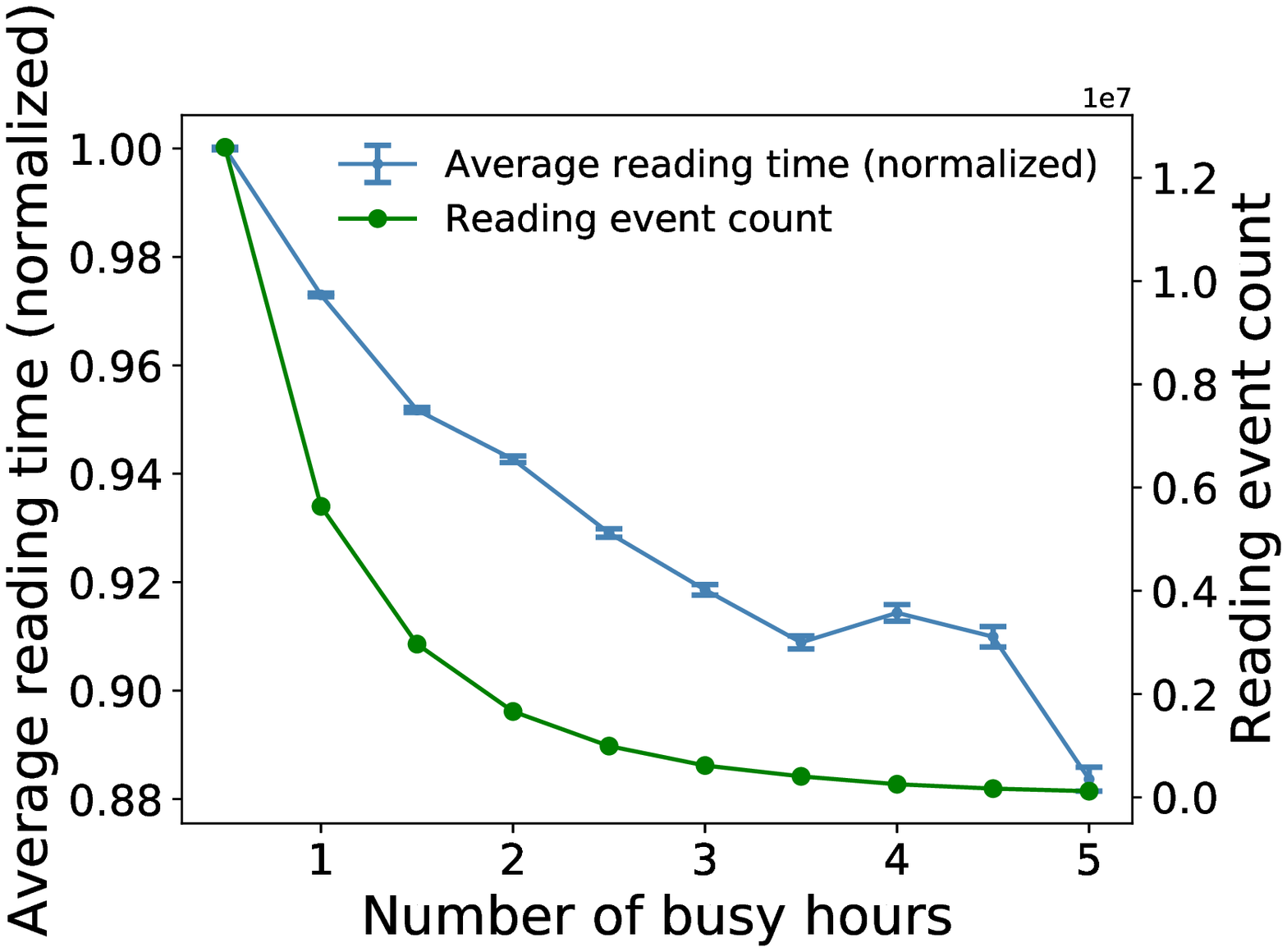}
    \caption{Reading time of desktop users by the number of meetings (top) and busy hours (bottom) in a day.} 
    \label{fig:busy}
\end{figure}

In Figure~\ref{fig:busy}, we consider the number of daily meetings, and the number of busy hours in the day as proxies for cognitive load of our desktop users. This is inspired by previous work by \citet{barley2011mail} that reported time spent at meetings as a source of stress at work. It turns out that more frequent meetings, and a larger number of busy hours in the day, are indeed correlated with observing fewer email reading events overall. That is, busy users read fewer emails, and go through those faster than average. We observed similar trends for our mobile users and hence exclude more details for brevity.

\paragraph*{Reading fatigue}
\label{sec:engagement}
Psychological research has shown that fatigue after mental work (e.g., proof reading) leads to a performance drop, such as reading speed and reaction time~\citep{ahsberg2000perceived}. But how does fatigue impact reading time in the email setting? 

\begin{figure}[tp]
        \centering  
         \includegraphics[width=\columnwidth]{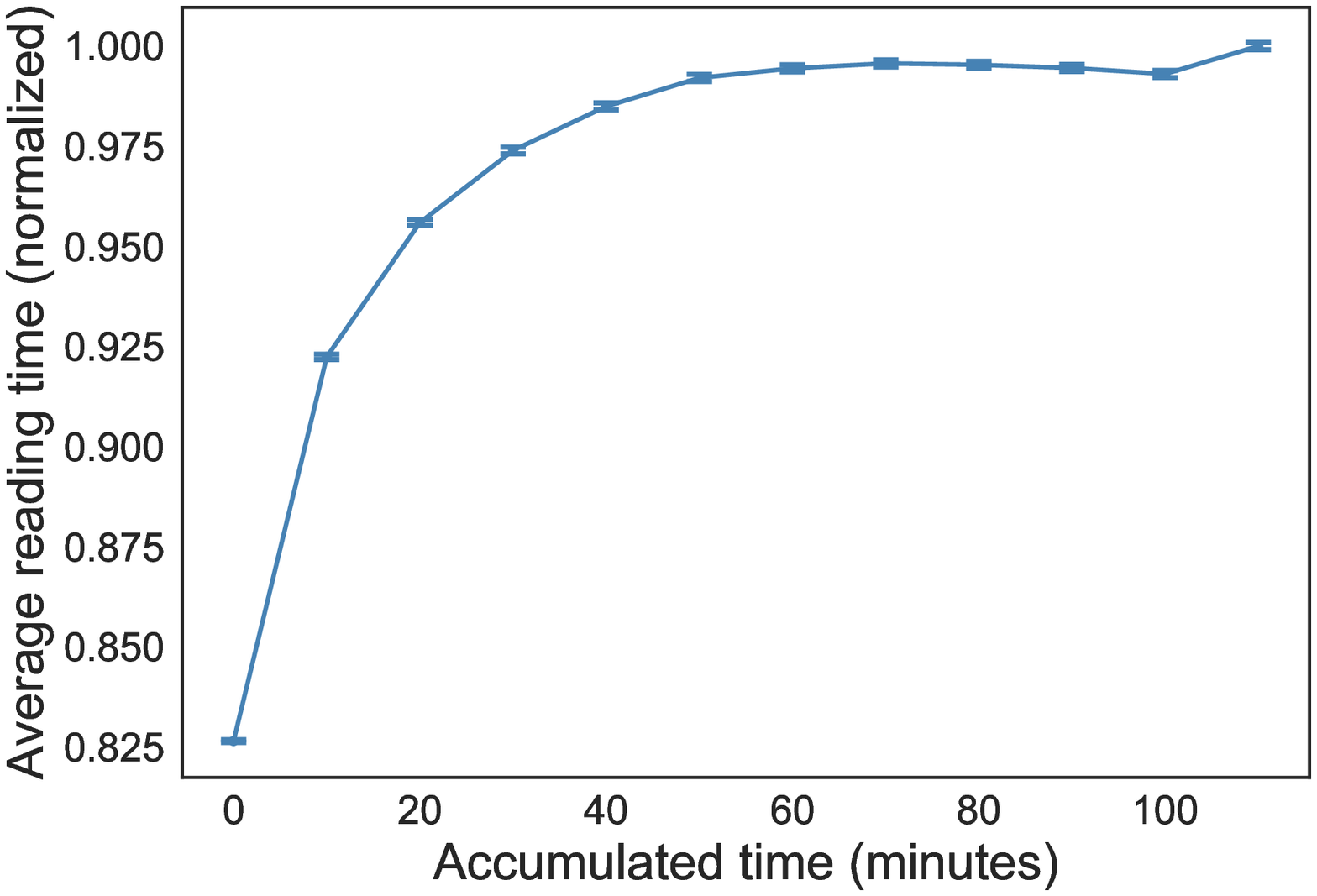}
         \caption{Average reading time conditional on the time spent on reading emails during the past 2 hours of user activity (proxy for fatigue).}
         \label{fig:fatigue}
\end{figure}

We use the accumulated time spent on reading emails in the past two hours of user activities as a proxy for measuring fatigue. The longer the accumulated time, the more we expect the user to be affected by fatigue. For each email reading event, we sum the accumulated reading time of the user in the past two hours prior to that event, and group these sums with a bin size of 10 minutes. The results can be found in Figure~\ref{fig:fatigue}. 
As accumulated reading time increases up until 60 minutes, the average reading time constantly grows. After that point the reading time does not change much. Although we cannot draw strong conclusions based on these observations in the absence of more information about the users, these trends \textit{may} suggest that fatigue prolongs reading time, and the effect is only up to a certain extent. Overall, our findings are consistent with the reported effect of fatigue in email settings by~\citet{ahsberg2000perceived}.

%\subsection{Device}
%\label{sec:device}

\paragraph*{User device}
We confine the scope of this part of study to mobile users because the device type information -- specifically, screen size details --  is only available to us in our mobile logs. Mobile devices have different screen sizes, which we hypothesize could impact the reading experience. In Table~\ref{table:device}, we group devices by their screen size and present their average reading time per email. Users on the smallest screen devices spend the longest reading time. This may be explained by the limited contents displayed on a small screen, which demands more efforts (scrolling, zooming) to read. The 9.7in iPads which have the largest screen size have the lowest reading time across all devices. Overall, the reading time is negatively correlated with screen size.%The smallest screen size belongs to iPhone SE which also holds the longest average email reading time in our logs.
% However, we do not observe significant differences of reading time among large screen devices.

\begin{table}[tp]
\centering
\caption{Reading time on different devices, ordered by device screen size.}
\label{table:device}
  \begin{tabular}{p{.1\columnwidth}p{.3\columnwidth}p{.25\columnwidth}p{.15\columnwidth}}
    \toprule
    \bf Screen size (inch)& \bf Average reading time (normalized)& \bf Sample device & \bf Percentage\\
    \midrule 
    4.0 &  1 & iPhone SE& 14.3\% \\
    4.7 & 0.89 & iPhone 7& 57.2\% \\
    5.5 & 0.89 & iPhone 7 Plus & 23.9\% \\
    7.9 & 0.90 & iPad mini 4 & 0.9\% \\
    9.7 & 0.82 & iPad Pro & 3.3\% \\
    %12.9 & 12.5 & iPad Pro (12.9 inches) & 0.3\% \\ %Traffic too small
  
    \bottomrule
    \end{tabular}
\end{table}

%% file: wsdm2018-xinyi-7.tex
% !TEX root = ./wsdm2018-xinyi.tex
\subsection{Reading and rereading}
\label{sec:rereading}
\begin{figure}[tp]
        \centering  
         \includegraphics[width=\columnwidth]{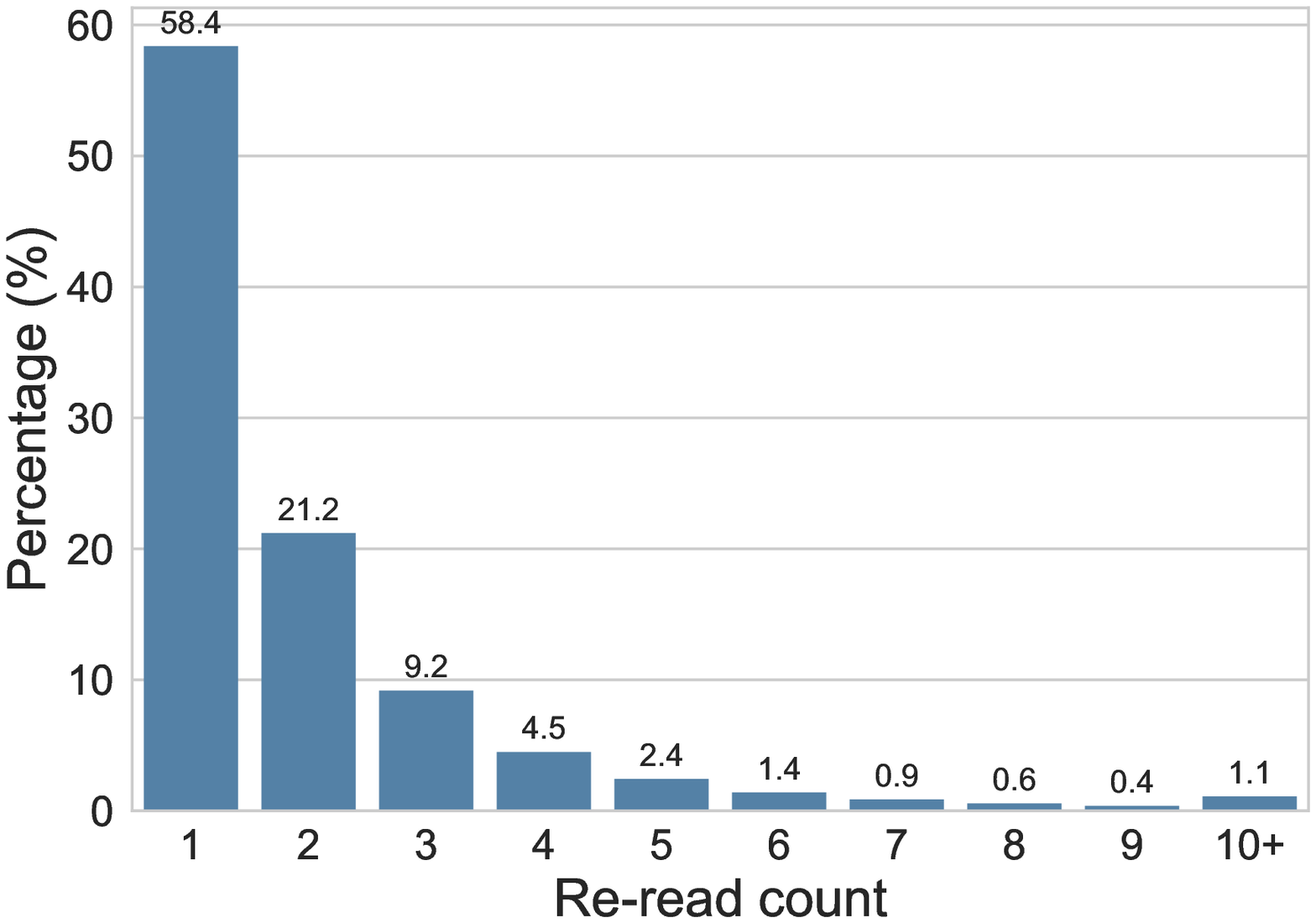}
         \caption{The distribution of email reread actions.}
         \label{fig:rereadcount}
\end{figure}

In this section we investigate the reading time of emails that have been read at least once before. We find that 33\% of unique email reading actions are actually \textit{rereads}, a significant portion that may seem surprising at first glance. However, as an interesting reference point, \citet{teevan2007information} reported that about 38.8\% of all web search queries are re-finding, which further underlines the scope of re-finding activities beyond email. It is worth noting that unlike Section~\ref{sec:method}, here if one email has say three recipients, it is considered as three unique emails for the purpose of computing reread statistics. 
Figure~\ref{fig:rereadcount} shows the distribution of reread counts for reread emails. We observe that in 58.4\% of the cases emails are reread once (read twice in total), while the majority of reread emails (95.7\%) are reread no more than 5 times.

Given the high frequency of rereading behavior observed in the logs, we are encouraged to extend our investigation further to compare reread actions for different email types, study the impact of previous reread counts on reading time, and analyze rereading cross-device. 

\begin{figure}[tp]
        \centering  
         \includegraphics[width=\columnwidth]{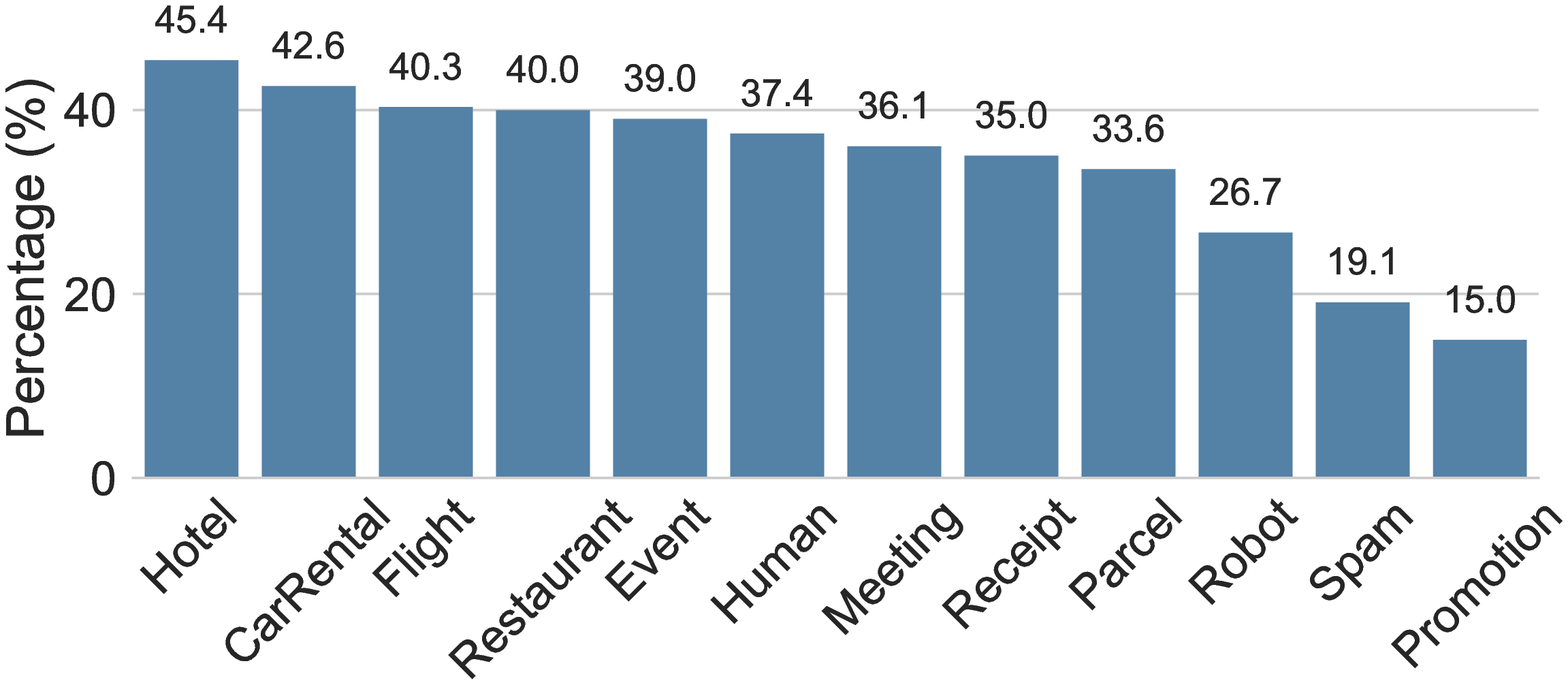}
         \caption{The percentage of emails that are reread across different classes of emails.}
         \label{fig:rereadPercDist}
\end{figure}

\paragraph*{Reread action across email types}
The results in Figure~\ref{fig:rereadPercDist} reveal that certain categories of emails (e.g. hotel, car rental and flights) are more likely to be reread. This may be explained by the way users deal with travel related emails, e.g. they may read it upon first delivery, but when they check in at the hotel or catch the flight they need to read it again for information. Human emails have a much higher rereading percentage (37.4\%) versus robot emails (26.7\%). While spam and promotion have the lowest rereading percentage (19.1\% and 15.0\%), they represent a noticeable portion in the emails that are read. One possible explanation is that during email triaging some users read spam or promotion, and they may flag or move some important emails to their inbox. Revisiting those later will be seen as rereading.

%\ms{dropping the reading time per email type stuff (commented)}
\if0
How does reading time change when an email is reread? We show the comparison between this pair of reading time in Figure~\ref{fig:rereadTimeDist}. For most email types, email reading time increases in rereading except for human, flight and receipt emails. Robot emails have bigger differences in reading time compared to human emails.

\begin{figure}[tp]
        \centering  
         \includegraphics[width=\columnwidth]{figs/readtime_1st_2nd_per_category.eps}
         \caption{Reading time comparison between 1st time read and reread per email type.\ms{This plot and its related text can potentially be dropped to save some time.}}
         \label{fig:rereadTimeDist}
\end{figure}
\fi

\begin{figure}[tp]
        \centering  
         \includegraphics[width=\columnwidth]{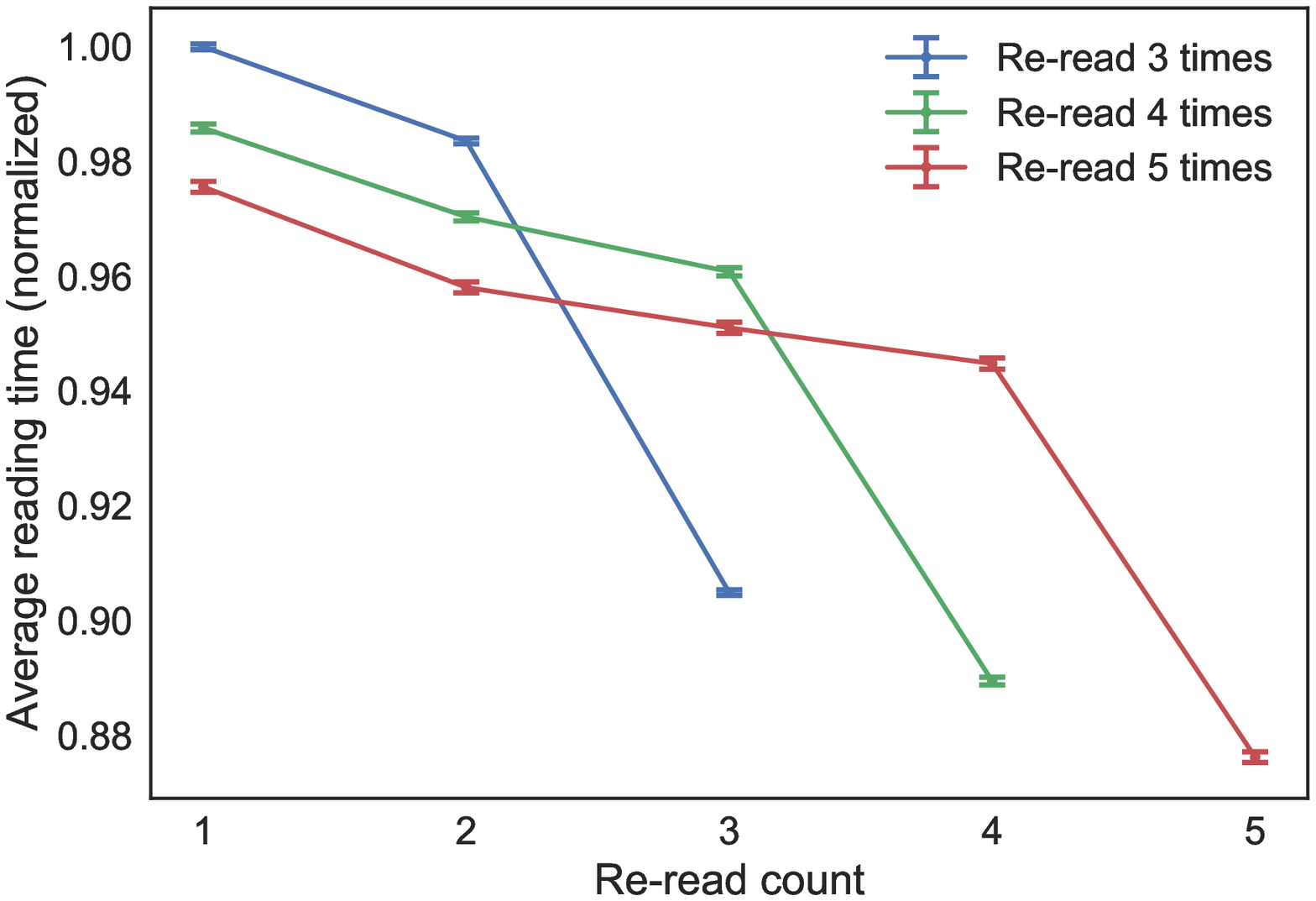}
         \caption{Average reading time of reread emails for different number of previous reads.} %\ms{please change reread to reread in the figure}}
         \label{fig:rereadTimeCount}
\end{figure}

\paragraph*{Reread count vs. reading time}
Earlier in Figure~\ref{fig:rereadcount} we described that 41.6\% of reread emails are reread more than once. How does reading time change, as users read certain emails over and over again? 
In Figure~\ref{fig:rereadTimeCount}, we look at how reading time changes by the reread count. We group the reread emails by the maximum number of times they have been reread from three to five. We observe that as emails are reread more often, users spend less time on reading when they have to go through them again, which probably can be explained by their increasing familiarity with the content. %Besides, when the email is read for the last time, it has significantly lower reading time compared to the previous reading event.
Another interesting finding is that emails that are reread 5 times are read more quickly than those reread 4 times, and those reread 4 times are read more quickly than those reread 3 times.

\paragraph*{Cross-device rereading}

With the increasing popularity of smart mobile devices, users can now easily switch to their mobile device to handle emails when they are away from their desktop. In this section we focus on emails that are reread across devices. We only include emails that are received during our sampling period and opened on more than one device. In total we have 587,953 emails and 67,440 users. Specifically, we are interested in cases where users read a new email on mobile first then switch to desktop to read it again, and vice versa.

One prominent characteristic of cross-device reading events is that 75.6\% of the emails are first read on mobile before being read on desktop. On the contrary, only 24.4\% of emails are first read on desktop, followed by mobile. One reason for this imbalance could be due to the convenient access to mobile devices, that enable users to get to their emails more easily and regularly. 
Another reason, could be that access to more information and easier typing on desktop encourages people to continue and finish the tasks they initially started on mobile, on desktop.

We also notice that when users switch from desktop to mobile, 29.5\% of the times the subsequent reading events happen within 30 minutes, while from mobile to desktop the percentage is much lower at 16.7\%. 

\begin{figure}[tp]
	\centering
        \begin{subfigure}[b]{\columnwidth}
        %\centering
        		\includegraphics[width=\linewidth]{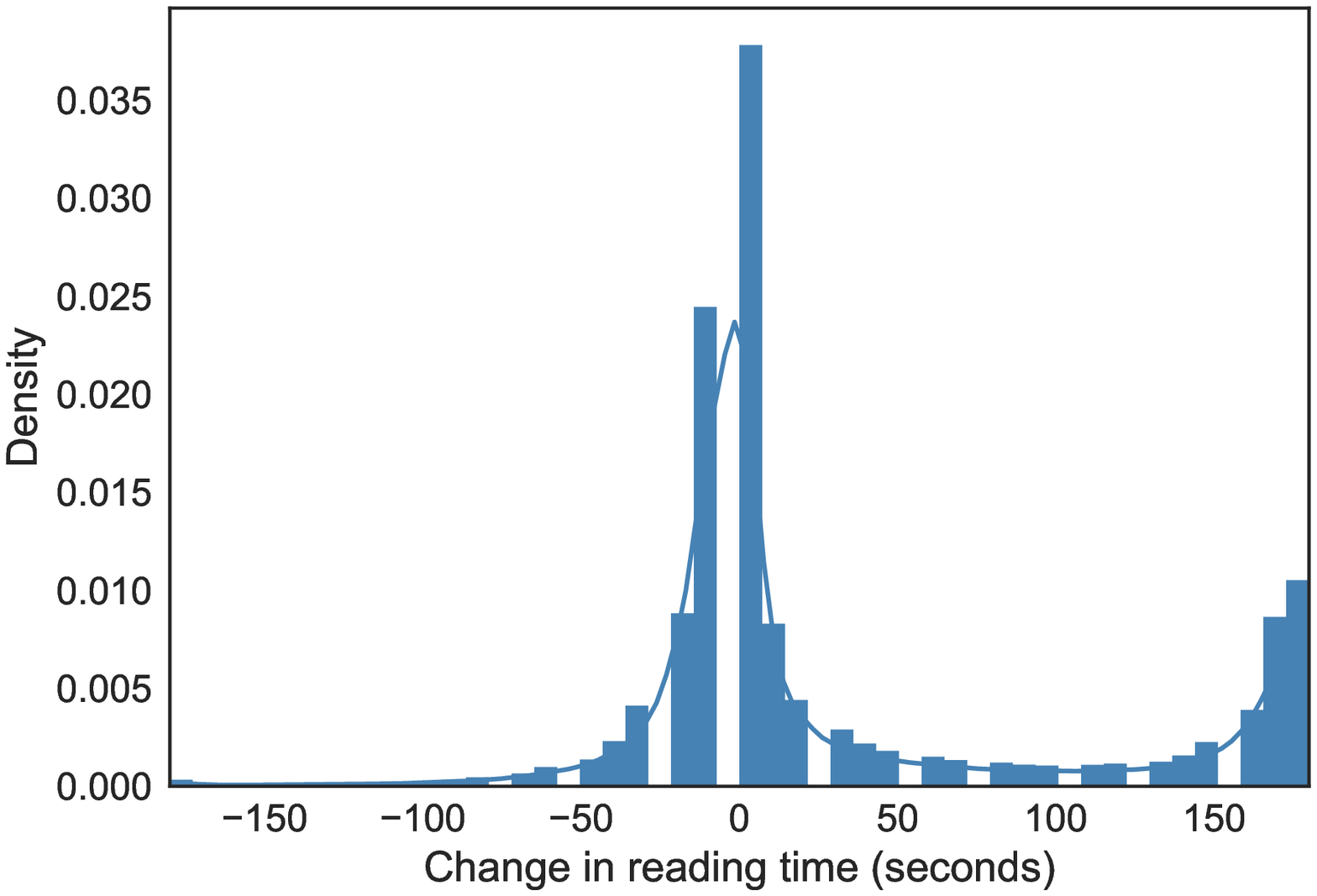}
       		\caption{Change in reading time from mobile to desktop.}
       		\label{fig:switchtime1}
        \end{subfigure}
        \hfill
       \begin{subfigure}[b]{\columnwidth}
        \centering
                \includegraphics[width=\linewidth]{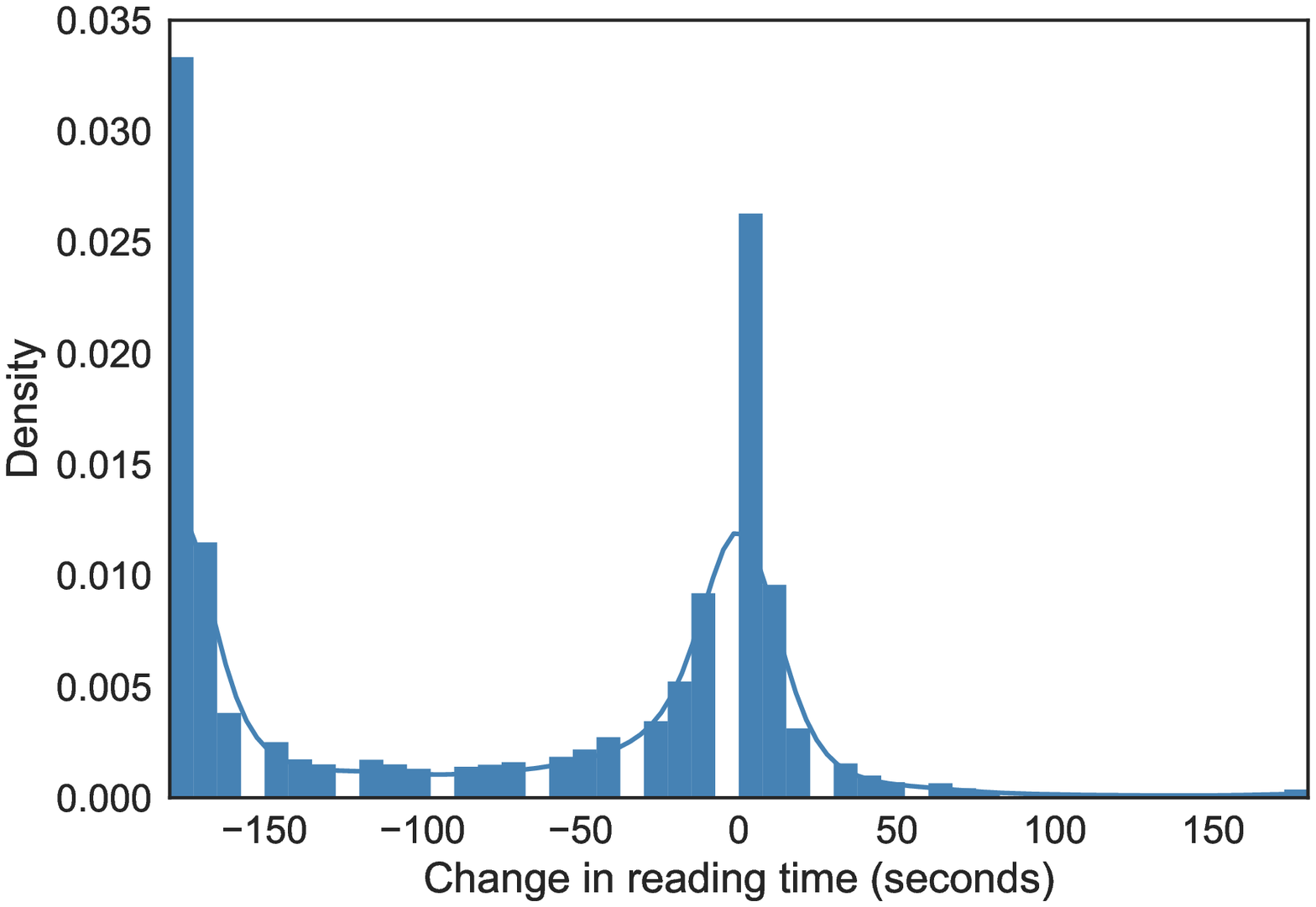}
                \caption{Change in reading time from desktop to mobile.\newline}
                \label{fig:switchtime2}
        \end{subfigure}
        
    \caption{Change in reading time when the user rereads the same email across platforms.} 
    \label{fig:switch}
\end{figure}

Finally, we explore how reading time changes after a user switches from one client to another. To this end, we measure how the reading time of an email first opened on mobile changes when the user opens it again on desktop and vice versa.
Figure~\ref{fig:switch} demonstrates the
histogram of changes\footnote{That is, reading time measured on the second client, minus the reading time recorded on the first client for the same email (in seconds).} in reading time for mobile to desktop switches on top, and for desktop to mobile switches at the bottom.
On both histograms, there is a large peak around 0, that suggests users spend roughly similar time when reading the same email across different platforms.
The right peak on the top histogram represents a large set of emails that are opened for the first time on mobile, and re-opened later on desktop with significantly longer reading time. This set is likely to include emails that the user has glanced through on mobile, but left to fully address on desktop where it is easier to type and access information. The left peak on the bottom histogram includes another interesting set of emails. These are emails that are opened on desktop for the first time and are reread on a mobile device, but with much shorter reading time on average. Perhaps these are emails that the users revisited for quick fact checks and referencing on mobile, when they had no access to their desktop client.

%% file: wsdm2018-xinyi-8.tex
% !TEX root = ./wsdm2018-xinyi.tex
\section{User Study}
\label{sec:userStudy}
%\ms{If the user study only confirms the observations from the logs and not addressing the gaps, we should consider dropping it.} \cj{Xinyi, can you point out which findings are only available in user study but not in the logs?} \xinyi{The user study provides proof for assumptions made in the observations. I'd say most our guesses turn out to be true in the user study. @Sue?}
%\sd{User studies or surveys can provide more information about "why" we see certain behaviors in the logs -- the "why" vs. the "what". E.g., we don't need a study to show that x\% of emails are re-visited since we know this using a much larger and more representative log-based sample.  However, the study can help us say something about "why" people are re-visiting.  Pulling out such insights will be especially interesting.}

User studies can provide more information about ``why" we see certain behaviors in the logs. Through a brief user study, we obtain qualitative explanations for some interesting observations, namely, what lead to long reading time on an email, why users conduct rereading, how and why users read emails across devices.

\noindent
\subsection{Methodology} 
Our user study consists of two phases: 1) screen recording; and 2) interview. We first record a one hour video on the user's computer screen, without interrupting the user's normal activities during work hours. The screen recording is one of the least intrusive ways to help us observe natural behavior on their working computer. 

%The recording is performed on desktop only, as all our participants use it as their main working platform. We use the screen recording function embedded in Office PowerPoint, because the software is readily available on all their desktops. The screen recording is conducted on the display where they use the email client. In case of multiple displays, only the one that shows the email client will be recorded. The recording output is saved as a media file for later review.

Then we conduct the interview a few hours later. We avoid interviewing immediately after recording so that the user will not hurry their work during the recording time. A one-on-one interview is conducted by playing back the recorded video and asking questions. The video helps the user to remember what he/she was doing during the recording. The interviewer also looks at the screen if permission is granted by the participant. Otherwise the interviewer sits back from a distance to avoid reading the contents on the screen. Specifically, the interview starts with general questions such as user's habits of reading emails, then moves into detailed questions on the user's interaction with emails during the recording. In total, 15 participants from an IT enterprise took part in this user study, including 9 men and 6 women. The demographics also include people at different ages (from 20s to 60s), and at different job positions (3 interns, 4 junior employees, 5 senior employees, and 3 senior managers). Afterwards, participants are awarded a 25 dollar coupon for online shopping.

\noindent
\subsection{Findings}
\textit{Long reading time.} In our sample, no one closes the email client during their work hours. All participants either keep it opened or minimized in the background. This can lead to excessive long reading time being recorded for the last opened email in a session, which partially explains the heavy tail for desktop reading time (the reading time longer than 3 minutes) in Figure~\ref{fig:overview}. Another reason that may explain the inflated reading time is participants' multi-tasking. Multitasking is frequent for the majority of participants and happens when it is needed to refer to the email contents to complete another task. However, neither of these 2 situations applies to mobile because users would usually close the mobile app after they use it, and they do not multi-task while reading emails on mobiles. This helps explain the much smaller tail of mobile reading time than that of desktop.

\textit{Rereading.} Rereading is common for all participants. It takes place either after an email search, or simply by browsing through emails that are read. We find two cases that frequently lead to rereading. The first is email triaging, especially for the senior employees and managers who may receive too many emails to finish reading at once. In this case, participants would flag emails or move them to certain folders after a quick skim, then read the emails again some time later. Another case is for difficult or long emails, that participants need to take several reading attempts to digest the content. This also includes the scenario when they first read the email on mobile, and continue processing it on desktop.

\textit{Cross-device reading.} 40\% of the participants report the use of the mobile client for their work email. They report using the mobile client only when they are away from the working environment, for instance morning at home, during transit or away for coffee breaks. For heavier reading tasks, participants would switch from the mobile client to the desktop. This could explain the reading time increase from mobile to desktop. In summary, the mobile client only serves as a complementary platform to the desktop client.

%% file: wsdm2018-xinyi-9.tex
% !TEX root = ./wsdm2018-xinyi.tex
\section{Discussion and Conclusions}
\label{sec:conclusion}
%\xinyi{single out applications in a separate section?}
%\xinyi{rewrote this section to highlight application scenarios}
This paper characterized in depth how people read their enterprise emails on desktop and mobile devices. We acknowledge that a limitation of this study is that direct applications are not provided, as the paper focuses on observational insights. However, the rich findings can open up directions for possible applications in email system design, as well as fostering research in email systems.

\textbf{Adaptable reading pane.} Email types and lengths affect reading time substantially. For instance, we found that people tend to spend more time reading human-authored emails, while ignoring spam or promotions. This may suggest, from users' perspectives, loading an entire promotion email to the reading pane is unnecessary, and the saved space could be utilized to support other ``smart'' options such as one-click unsubscribe.  
%We also found that the number of times users spend reading an email is usually negatively correlated with the number of recipients on that email. However, time spent on reading single-recipient emails appears to be the shortest. 

\textbf{Contextual inbox.} People tend to be more active reading on desktops during morning and noon hours, whereas on mobile devices reading time increases from evenings to midnights. As expected, our temporal analysis suggests that attention is more focused on work-related communications on weekdays, and on travel activities during the weekends. These findings can help us build a contextual inbox. For instance, reducing pop-up notifications for receipt confirmation emails can potentially help users stay focused in a meeting. Likewise, for important/urgent emails that are delivered at night, auto-replies to senders and reminders to recipients' mobile phones may help reduce tension and response latency.     

\textbf{Email assistant.} We find that people with busier calendar schedules may read fewer emails and process those faster than average. Similar to the fatigue effect identified in psychological research conducted by~\cite{ahsberg2000perceived}, we find that the longer accumulated time users spend reading in the past two hours, the slower they may become in terms of processing new emails. In such cases, email clients can track down things that need to be completed, highlight items that are skipped due to a lack of concentration time, or even auto-complete them (e.g., schedule a follow-up meeting per discussed), which may alleviate users' burdens from busy days.

\textbf{Cross-platform rereading support.}
Users also reread emails across platforms, where 76\% of cross-platform rereadings happen first on mobile then on desktop, and 24\% vice versa. For the former case, users tend to continue heavy tasks on desktop. The system can assist users' rereading activity by remembering the last-read position and help them continue processing the email. For the latter, since users spend significantly less time when rereading on mobile (e.g. fact checking), summarization and highlighting email contents would save user efforts and improve their efficiency.

%We found revisiting previously read emails is a common phenomenon. Users prefer to reread human emails (37.4\%) over robot emails (26.7\%), and spend less time when they reread the same email multiple times. 

Our log analysis has painted a rich picture of reading time on emails in general. A user study in an IT enterprise also served as a sanity check for the observations. Further, it would be interesting to investigate how the nature of the business affects the email reading behavior (e.g., a production-based company will possibly be very different to a government organization). Although this was not covered in the log analysis due to privacy protection on user identity (we do not have access to email addresses), conducting pop-up surveys as in \cite{Kim:2017:UMS:3077136.3080837} can provide large-scale supportive evidence that helps complement our log analysis and user study. We also discussed several ways how these findings could be used. The action ``reading'' is shared across different email-related scenarios. If we understand reading time for a user query and the corresponding search success, can we infer and adjust our understanding of, for example, user reading time on an auto-generated reminder or meeting invitation? We leave these interesting questions to our future work.